\documentclass[aps,prx,reprint,preprintnumbers,superscriptaddress,nofootinbib,longbibliography,floatfix]{revtex4-2}
\pdfoutput=1
\usepackage{rotating}
\usepackage{array}
\usepackage{amsmath}
\usepackage[normalem]{ulem}
\usepackage{slashed}
\usepackage{booktabs}
\usepackage[pdftex,table]{xcolor}
\usepackage{units}
\usepackage{xfrac}
\usepackage{mathtools}
\usepackage{empheq}
\usepackage[]{units}
\usepackage{multirow}
\usepackage{amssymb}
\usepackage{url}
\usepackage{comment}
\usepackage{physics}
\usepackage{color,soul}
\usepackage{bbm}
\usepackage[caption=false]{subfig}
\usepackage{adjustbox}
\usepackage[T1]{fontenc}
\usepackage{xspace}

\usepackage{graphicx,caption,subcaption,tikz}
\usetikzlibrary{backgrounds}

\usetikzlibrary{calc}
\DeclareCaptionFormat{overlay}{\gdef\capoverlay{#1#2#3\par}}
\DeclareCaptionStyle{overlay}{format=overlay}
\DeclareCaptionFormat{suboverlay}{\gdef\subcapoverlay{(\thesubfigure) #3\par}}
\DeclareCaptionStyle{suboverlay}{format=suboverlay}

\usepackage{hyperref}
\hypersetup{
  colorlinks=true,
  citecolor=blue,
  linkcolor=blue,
  urlcolor=blue
}

\newcommand{\ours}{\textsc{Parnassus}\xspace}

\begin{document}

\title{\ours: An Automated Approach to Accurate, Precise,\\ and Fast Detector Simulation and Reconstruction}

\author{Etienne Dreyer}
\email{etienne.dreyer@weizmann.ac.il}
\affiliation{Weizmann Institute of Science, Rehovot, Israel}

\author{Eilam Gross}
\email{eilam.work@gmail.com}
\affiliation{Weizmann Institute of Science, Rehovot, Israel}

\author{Dmitrii Kobylianskii}
\email{dmitry.kobylyansky@weizmann.ac.il}
\affiliation{Weizmann Institute of Science, Rehovot, Israel}

\author{Vinicius Mikuni}
\email{vmikuni@lbl.gov}
\affiliation{Lawrence Berkeley National Laboratory, Berkeley, CA 94720, USA}

\author{Benjamin Nachman}
\email{bpnachman@lbl.gov}
\affiliation{Lawrence Berkeley National Laboratory, Berkeley, CA 94720, USA}

\author{Nathalie Soybelman}
\email{nathalie.soybelman@weizmann.ac.il}
\affiliation{Weizmann Institute of Science, Rehovot, Israel}

\begin{abstract}
Detector simulation and reconstruction are a significant computational bottleneck in particle physics. We develop Particle-flow Neural Assisted Simulations (\ours) to address this challenge. Our deep learning model takes as input a point cloud (particles impinging on a detector) and produces a point cloud (reconstructed particles). By combining detector simulations and reconstruction into one step, we aim to minimize resource utilization and enable fast surrogate models suitable for application both inside and outside large collaborations. We demonstrate this approach using a publicly available dataset of jets passed through the full simulation and reconstruction pipeline of the CMS experiment. We show that \ours accurately mimics the CMS particle flow algorithm on the (statistically) same events it was trained on and can generalize to jet momentum and type outside of the training distribution.

\end{abstract}

\maketitle


\section{Introduction}
\label{sec:intro}

Synthetic datasets are essential for data analysis in all areas of particle physics. Such datasets are an accurate and precise representation of nature; they bridge the fundamental theory to observable quantities. The full science program of a given particle physics experiment requires many synthetic datasets, with many variations on fundamental (Standard Model and beyond), phenomenological, and instrumental parameters. This is a significant computational bottleneck that is particularly acute for the ATLAS and CMS experiments at the Large Hadron Collider (LHC), where simulation and reconstruction of synthetic data during the high-luminosity era will require more computing resources than real data processing~\cite{CERN-LHCC-2022-005,Software:2815292}. 

The community's solution to this challenge is fast simulation. Instead of modeling all of the detailed physical aspects of the detector response, fast simulations approximate some or all aspects that a full simulation tool based on frameworks like \textsc{Geant}4~\cite{GEANT4:2002zbu,Allison:2006ve,Allison:2016lfl} would include. ATLAS~\cite{ATLAS:2010arf,ATLAS:2010bfa,ATLAS:2021pzo,ATLAS:2022jhk}, CMS~\cite{Abdullin:2011zz,Giammanco:2014bza,Sekmen:2016iql}, 
and other large collaborations spend years designing fast detector simulations that produce the same output as a full detector simulation. These outputs are then processed with the standard reconstruction. Machine learning is playing a growing role in fast detector simulations, with the ability to automatically train end-to-end from particle input to detector readout, modeling all of the complex correlations. This includes Generative Adversarial Networks (GANs)~\cite{GANs}~\cite{GanPhys2,GanPhys3}, Variational Autoencoders~\cite{VAEs}~\cite{ATL-SOFT-PUB-2018-001}, Normalizing Flows (NFs)~\cite{NFs}~\cite{caloflow1}, and Diffusion Models~\cite{scoremodels}~\cite{mikuni:caloscore} (only first papers cited; see Ref.~\cite{Feickert:2021ajf,Adelmann:2022ozp,Hashemi:2023rgo} for reviews). The output of the fast detector simulations is very high-dimensional, set by the fine granularity of modern particle detectors. However, most users do not make use of this high-dimensional data - they only interact with the reduced data following reconstruction. For most applications, this process is nearly lossless in the sense that the reconstructed particle four-vectors and particle type are a statistically sufficient representation of their constituent readout elements. Furthermore, the simulation and reconstruction of synthetic data each require about the same resources - solving one is not sufficient to address the full computational challenge~\cite{CERN-LHCC-2022-005,Software:2815292}.

Most studies in particle phenomenology and innovative algorithms for collision data are first developed outside of large collaborations. Researchers in this context typically lack the computational resources and expert knowledge required to deploy the simulation software of the experiments they are interested in. Instead, they rely on user-friendly tools with simplified simulation and reconstruction algorithms~\cite{deFavereau:2013fsa,Selvaggi:2014mya,Mertens:2015kba,pgs,Buckley:2019stt,Araz:2020lnp,Vohra:2022hzq}. The most widely-used tool is \textsc{Delphes}~\cite{deFavereau:2013fsa,Selvaggi:2014mya,Mertens:2015kba}, which is based on smearing the properties of truth particles combined with a simplified reconstruction model. The smearing functions are manually added per particle type, energy, and direction using parameters extracted from public performance plots of experimental collaborations. It requires a dedicated effort to keep these parameters aligned with the evolving conditions at current experiments, resulting in lag, especially when performance inputs are not yet publicly available. Since the detector geometry and reconstruction algorithms inside these tools are highly simplified, even if the smearing functions are exactly correct, the resulting synthetic datasets do not exactly mimic those of the target experiment.

Our goal is to create an end-to-end fast simulation paradigm that unifies internal and external users, can be automatically tuned, and includes both generation and reconstruction. The inputs to this program are stable particles entering the detector, and the outputs are reconstructed particles. This is enabled through a deep generative model creating a point cloud (reconstructed particles) conditioned on another point cloud (detector-stable particles). Since it is based on a neural network, the entire framework is written in Python with few dependencies and automatically compatible with Graphical Processing Units (GPUs). The output dimension is the same order of magnitude as the input dimension and so this tool should be computationally accessible to all users.  Our vision is that models could be trained on full-simulation samples from experiments and then published so that the broader research community has access to the most accurate and precise versions for their studies.  Generic detector simulations like \textsc{Cocoa}~\cite{Charkin-Gorbulin:2023zpz} could also have fast simulations published for general consumption.  We call our approach \ours\footnote{The widely-used event generator \textsc{Pythia}~\cite{Sjostrand:2006za} is a reference to the oracle at Delphi; presumably the \textsc{Delphes} detector simulation is accordingly an ancient-Greek reference.  Delphi sits on top of Mount Parnassus, which motivated our name.}: a \underline{Par}ticle-flow \underline{N}eural-\underline{as}sisted \underline{S}im\underline{u}lation\underline{s}. In this paper, we demonstrate the \ours setup with high-energy hadronic jets. Subsequent work will be required to scale up this approach to full collision events and to provide a user-friendly interface for the \ours framework.

There have been a number of important and foundational studies related to our proposal. In addition to detector simulation,
machine learning has also been extensively studied for particle reconstruction~\cite{Pata:2021oez,Qasim:2022rww,DiBello:2022iwf,Kahn_2022,GarciaPardinas:2023pmx,Pata:2023rhh}, with some models already in production within experiments~\cite{ATLAS:2021pzo,ATLAS:2022jhk,ATL-PHYS-PUB-2022-040,Mokhtar:2023fzl}.  A number of point cloud generative models have been proposed for detector simulations~\cite{caloclouds,Buhmann:2023bwk,Acosta:2023zik,Buhmann:2023kdg,Schnake:2024mip,Kobylianskii:2024ijw}, and jet generation~\cite{Kansal:2021cqp,Kach:2022qnf,Buhmann:2023pmh,Buhmann:2023zgc,Leigh:2023toe,Mikuni:2023dvk,Kach:2023rqw,Leigh:2023zle}, usually with a small number of high-level conditional inputs (high-dimensional conditional inputs were studied in Ref.~\cite{Shmakov:2024gkd}). The closest previous research to our proposal are Refs.~\cite{2704573,Vaselli:2024vrx} and Ref.~\cite{fastsim}. The former proposes a conditional generative model to transform single particle-level objects into single detector-level outputs while the latter only considers charged particles and a simplified smearing. The object-level studies represent a complementary approach where the inputs/outputs include jets instead of individual hadrons. While the jet-based approach will be effective for many applications, our particle-based framework is generic\footnote{While we do not include displaced vertices in this study, we plan to include it in future versions of \ours; this is a technical and not conceptual addition.} in the sense that any observable (including any type of jet and any jet substructure) can be computed post-hoc. Furthermore, the detector response is nearly universal at the level of particles and so by modeling individual particles, the detector effects should be able to be as accurate as possible. Reconstruction effects, on the other hand, have a non-trivial dependence on the local environment of a given particle, such as in jets. To capture such effects, the predictions of \ours are conditioned on a transformer encoding of the truth jet constituents' features. Our choice of neural network model is informed by the companion paper in Ref.~\cite{Kobylianskii:2024sup}.

This paper is organized as follows. Section~\ref{sec:data} introduces the generator-level and detector-level synthetic samples used for the numerical studies. The machine learning methods are described in Sec.~\ref{sec:methods} and results are presented in Sec.~\ref{sec:results}. The paper ends with conclusions and outlook in Sec.~\ref{sec:conclusions}.

\section{Datasets}
\label{sec:data}

In order to demonstrate \ours, we take jets of generator-level particles and attempt to predict the corresponding sets of reconstructed particles after applying the full CMS simulation and reconstruction chain. Our goal is to show that we can reproduce this target far more accurately than \textsc{Delphes} configured with the appropriate CMS run card.

We use the CMS 2011 Jet Primary dataset \cite{CMS_JetPrim}, which was reprocessed into the MIT Open Data (MOD) format for easy access~\cite{komiske_exploring_2020}. The high-energy quark-gluon scattering in these events was generated by \textsc{Pythia} 6.4.25 \cite{Sjostrand:2006za} with a detector simulation based on \textsc{Geant}4 \cite{GEANT4:2002zbu} and using particle-flow reconstruction~\cite{CMS:2017yfk}. Both the particle-flow candidates (PFCs) and the generator-level particles are clustered into jets using the anti-$k_t$ algorithm~\cite{Cacciari:2005hq,Cacciari:2011ma,Cacciari:2008gp} with a radius parameter of $R=0.5$. We consider particles inside the two highest $p_T$ jets in each event, with the two jets treated separately.
Each generator-level particle and PFC is characterized by its transverse momentum, direction, and charge ($p_T, \eta, \phi, |q|$). For a comparison, we converted the stable final-state particles from \textsc{Pythia} into the \textsc{HepMC}3 format~\cite{Buckley_2021} and simulated them using \textsc{Delphes} configured for CMS.
\begin{table}[b]
    \centering
    \begin{tabular}{cccc}
        \toprule
         $p_T^{\text{min}}$ - $p_T^{\text{max}}$ [GeV] & Type & Training & Testing\\\hline
         470 - 600 & Out-of-distribution &  & $\checkmark$ \\
         600 - 800 & Out-of-distribution &  & $\checkmark$ \\
         800 - 1000 & In-distribution & $\checkmark$ & $\checkmark$ \\
         1000 - 1400 & In-distribution & $\checkmark$ & $\checkmark$ \\
         1400 - 1800 & Out-of-distribution &  & $\checkmark$ \\
         1800 - $\infty$ & Out-of-distribution &  & $\checkmark$ \\
         \bottomrule
    \end{tabular}
    \caption{Information about used MC event samples provided by MOD dataset~\cite{MOD_Jet470,MOD_Jet600,MOD_Jet800,MOD_Jet1000,MOD_Jet1400,MOD_Jet1800}. Parton-level $p_T$ ranges and usage are shown. As shorthand, the samples are referred to as JX where X is $p_T^\text{min}$ in GeV.}
    \label{tab:samples}
\end{table}
To mitigate tracking inefficiencies and improve momentum reconstruction, we follow the Refs.~\cite{komiske_exploring_2020, Tripathee_2017} and require $p_T > 1$ GeV.

We used a total of six datasets split into different parton $p_T$ ranges, with two dedicated to model training and in-distribution performance evaluation, and the remaining four for out-of-distribution checks (see \autoref{tab:samples}). Each training dataset consisted of 1M jets, while each testing dataset comprised 200k jets, resulting in a total of 2M jets for training and 1.2M jets for testing.

The effect of multiple primary vertices per event (pileup) is included in the CMS samples at the level of detector response, but the generator-level pileup particles are absent from the truth record. To add pileup in \textsc{Delphes}, we extracted 40k truth events from the corresponding CMS minimum-bias samples \cite{CMS_MinBias}. These events are used to create the pileup file in the run configuration \texttt{delphes\_card\_CMS\_PileUp.tcl}, with the mean number of pileup vertices per event set to 6.35, as measured in a Poisson fit to the distribution in the CMS Jet Primary sample. Since the longitudinal coordinate of the hard scatter vertex is not recorded in the MOD dataset, we fix it at the origin, whereas pileup vertices are spread out according to the default CMS run configuration. Our \textsc{Delphes} results use the PFC collection incorporating all tracks (i.e., without any pileup subtraction). Similarly, no pileup subtraction is applied to the set of PFCs from CMS. Finally, the anti-$k_t$ algorithm ($R=0.5$) is run on the \textsc{Delphes} PFCs, and only constituents of the leading jet are kept.

\section{Methods}
\label{sec:methods}
\subsection{Continuous normalizing flows}
Continuous normalizing flows (CNF)~\cite{chen2019neural} are generative models that define a mapping between samples $x_0$ from a base distribution $p_0$ to samples $x_1$ from a target distribution $p_1$ in terms of an ordinary differential equation (ODE):
\begin{equation}
    \mathrm{d}x = v_{\theta}(t, x) \mathrm{d}t,
\end{equation}
where the vector field $v_{\theta}$ is modeled using a neural network and $t\in[0, 1]$. Recent work by Lipman et al.~\cite{lipman_flow_2023} proposed a Flow Matching (FM) objective, enabling the learning of this vector field through a regression task:
\begin{equation}
    \mathcal{L}_{\text{FM}}(\theta) = \mathbb{E}_{t,p_t(x)}||v_{\theta}(t, x) - u_t(x)||^2\,.
\end{equation}
The function $u_t$ generates the desired probability path between $p_0$ and $p_1$, but is generally not available in closed form. This problem can be solved by defining the Conditional Flow Matching (CFM) objective~\cite{lipman_flow_2023,tong2024improving}, where $u_t$ and the probability path $p_t$ can be constructed in a sample-conditional manner, resulting in:
\begin{equation}
    \mathcal{L}_{\text{CFM}}(\theta) = \mathbb{E}_{t, q(z),p_t(x|z)}||v_{\theta}(t, x) - u_t(x|z)||^2,
    \label{eq:cfm_loss}
\end{equation}
where $t \in [0, 1], z \sim q(z), x \sim p_t(x | z)$. Optimizing the CFM objective is equivalent to optimizing the FM objective. Our implementation uses the modified CFM formulation introduced in~\cite{lipman_flow_2023}. By identifying the condition $z$ with the single sample $x_1$ (for our case, a set of PFCs in the jet), we define:
\begin{equation}
    \begin{aligned}
        p_t(x|z) &= \mathcal{N}(x | tx_1, (t\sigma-t+1)^2)\,,\\
        u_t(x|z) &= \frac{x_1 - (1 - \sigma)x}{1 - (1-\sigma)t}\,,
    \end{aligned}
    \label{eq:prob_path}
\end{equation}
which represents a straight path between a standard normal distribution and a Gaussian distribution centered at $x_1$ with standard deviation $\sigma$, which we took to be equal to $10^{-4}$. 
We follow Ref.~\cite{dax_flow_2023} and sample $t$ from a power-law distribution $p(t) \propto t^{\frac{1}{1 + \alpha}}, \ t\in [0, 1]$. Setting $\alpha = 0$ results in a uniform distribution, whereas $\alpha > 0$ assigns greater weight to probability paths corresponding to larger $t$ values. Empirically, we found that $\alpha = 2$ yields optimal performance.

\subsection{Architecture description}
We employ a transformer-based neural network~\cite{DBLP:journals/corr/VaswaniSPUJGKP17} to parameterize $v_{t}$. The network receives inputs consisting of a fixed-length set of PFCs with $\log p_T$, $\eta$, and $\phi$ sampled according to \autoref{eq:prob_path}, as well as a set of truth particles with $\log p_T$, $\eta$, $\phi$, and $|q|$, along with event scaling information serving as global features. Before inputting the particles into the neural network, we order them\footnote{Each level is marginally permutation invariant, but since there is not a 1-1 match between levels, implementing joint permutation is non-trivial. This is something that could improve fidelity in future versions.} based on $p_T$ and apply a relative scaling. On a per-jet basis, we standardize $\log p_T$, $\eta$, and $\phi$ using the mean and standard deviations derived from the truth particles. The resulting scaling parameters for $\log p_T$, $\eta$, and $\phi$ serve as global features, ensuring the preservation of information regarding the absolute jet properties.

\begin{figure}[b]
    \centering
    \includegraphics[width=1\linewidth]{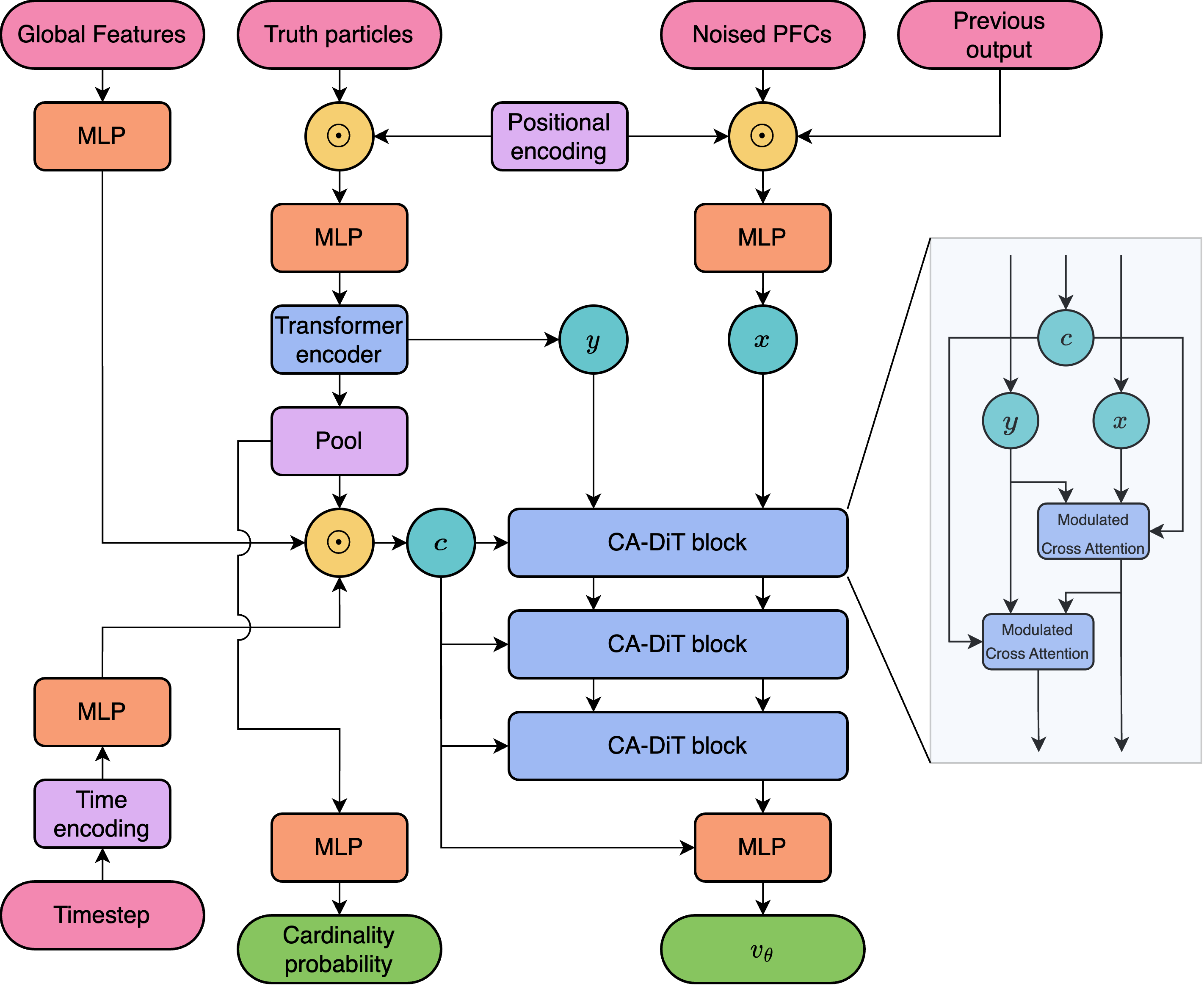}
    \caption{\textbf{Model architecture.} Concatenation is indicated by $\odot$.}
    \label{fig:model_scheme}
\end{figure}

The model architecture is illustrated in \autoref{fig:model_scheme}. Initially, sine positional encodings are added to the features of the noised PFCs and truth particles, followed by embedding them into the same hidden dimension using separate multi-layer perceptrons (MLPs). Subsequently, the representation of truth particles is updated via a transformer encoder to capture inter-particle dependencies. As shown in \autoref{fig:model_scheme}, the updated representations of PFCs and truth particles are then passed through three CA-DiT blocks. Each block updates the representations of both PFCs and truth particles using cross-attention along with adaptive layer normalization (adaLN-zero) to incorporate timestep and global information, as proposed by Peebles et al.~\cite{peebles_scalable_2023}. The inputs to the adaLN-zero mechanism
include sine encoding of the current timestep transformed with MLP, global scaling features, and the pooled representation of truth particles.
This pooled representation is also employed to predict the categorical probability of PFCs cardinality via a separate MLP, using a weighted Cross Entropy Loss term in addition to the CFM loss (\autoref{eq:cfm_loss}). The total loss is formulated as:
\begin{equation}
    \mathcal{L} = \mathcal{L}_{\text{CFM}} + 0.1\mathcal{L}_{\text{CE}}\,.
\end{equation}

To enhance sampling quality, we employ self-conditioning, a technique proposed in Ref.~\cite{chen_analog_2023}. This involves concatenating $v_t$ with the previously estimated $\tilde{v}$. During training, we introduce randomness by setting $\tilde{v} = 0$ with a probability of $p = 0.5$, effectively reverting to the model without self-conditioning. For the remaining instances, we first estimate $\tilde{v}$ and then utilize it for self-conditioning with detached gradients. During the inference, $\tilde{v}$ is always the neural network output from the previous timestep for $t>0$.

As we need to model a variable cardinality of the produced particle set, we divide the inference process into two stages.
Initially, the network predicts the probability vector for each potential cardinality based on the truth particle set. The number of generated PFCs is then sampled from a Multinomial distribution utilizing this vector. Subsequently, all PFCs are initialized with Gaussian noise.
Then, we employ a 4th-order PNDM method~\cite{liu_pseudo_2022} in conjunction with our network to iteratively predict the PFC features from this noise.

The network is implemented with the Pytorch~\cite{pytorch_my} package; hyperparameters are shown in~\autoref{tab:hyperparams}.
\begin{table}[h]
    \centering
    \begin{tabular*}{0.9\linewidth}{@{\extracolsep{\fill}}lr}
    \toprule
    \multicolumn{2}{@{}l}{\textbf{Hyperparameters}}\\\hline
    Batch size & 100\\
    Optimizer & AdamW~\cite{loshchilov2019decoupled}\\
    Weight decay & 0.01\\
    Learning rate & $10^{-4}$\\
    \# of epochs & 100\\
    \# of time steps & 25\\
    Gradient norm clip & 1.0\\
    Max output particles & 200\\
    Trainable parameters & 2,900,300 \\
    \bottomrule
    \end{tabular*}
    \caption{Network hyperparameters of the \ours model.}
    \label{tab:hyperparams}
\end{table}
\section{Results}
\label{sec:results}

The feature space we are generating is too big to examine holistically. Therefore, we consider a number of observables that are representative of the overall performance. To investigate the emulation of PFC kinematic features, we match PFCs to truth particles using Hungarian matching algorithm~\cite{HUNGARIAN}, where $\Delta R$ is used as a cost function. Since there is not a one-to-one correspondence between truth particles and PFCs, this matching is only approximate. At the level of entire jets, there is no matching ambiguity. In addition to probing kinematic properties of the constituents and the jets, we also explore a number of jet substructure quantities. These include the jet mass, the two-prong jet substructure taggers $C_2$~\cite{Larkoski:2013eya} and $D_2$~\cite{Larkoski:2014gra} (lower values are more two-prong like) as well as a state-of-the-art quark/gluon transformer-based tagger~\cite{Mikuni:2024qsr} trained on an independent dataset from the CMS Open Simulation.

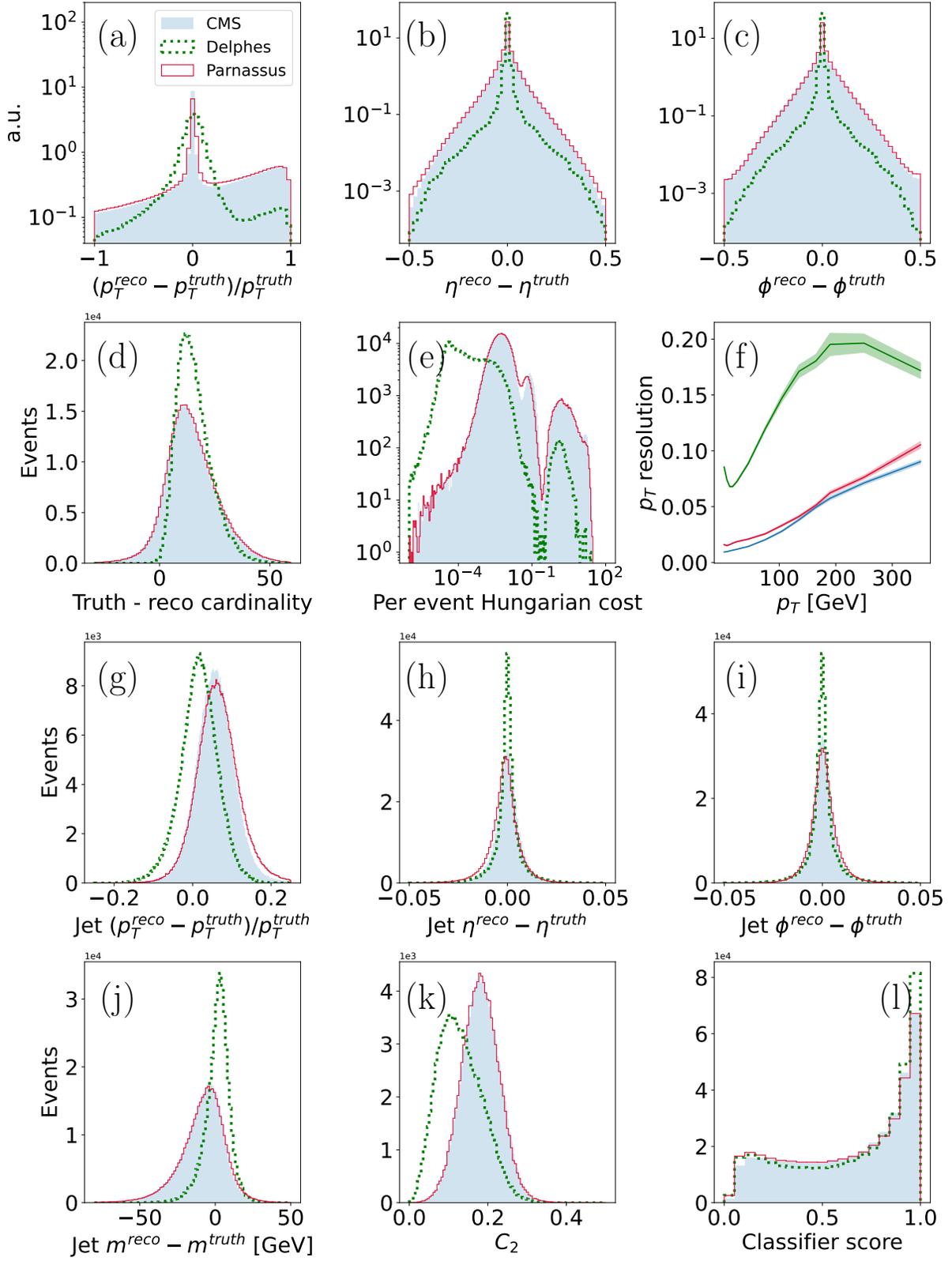
\begin{figure*}
    \centering
    \begin{tikzpicture}
    \path (0,0) node [red]                    {A};
     \node [inner sep=0] (image) {\includegraphics[width=0.9\textwidth]{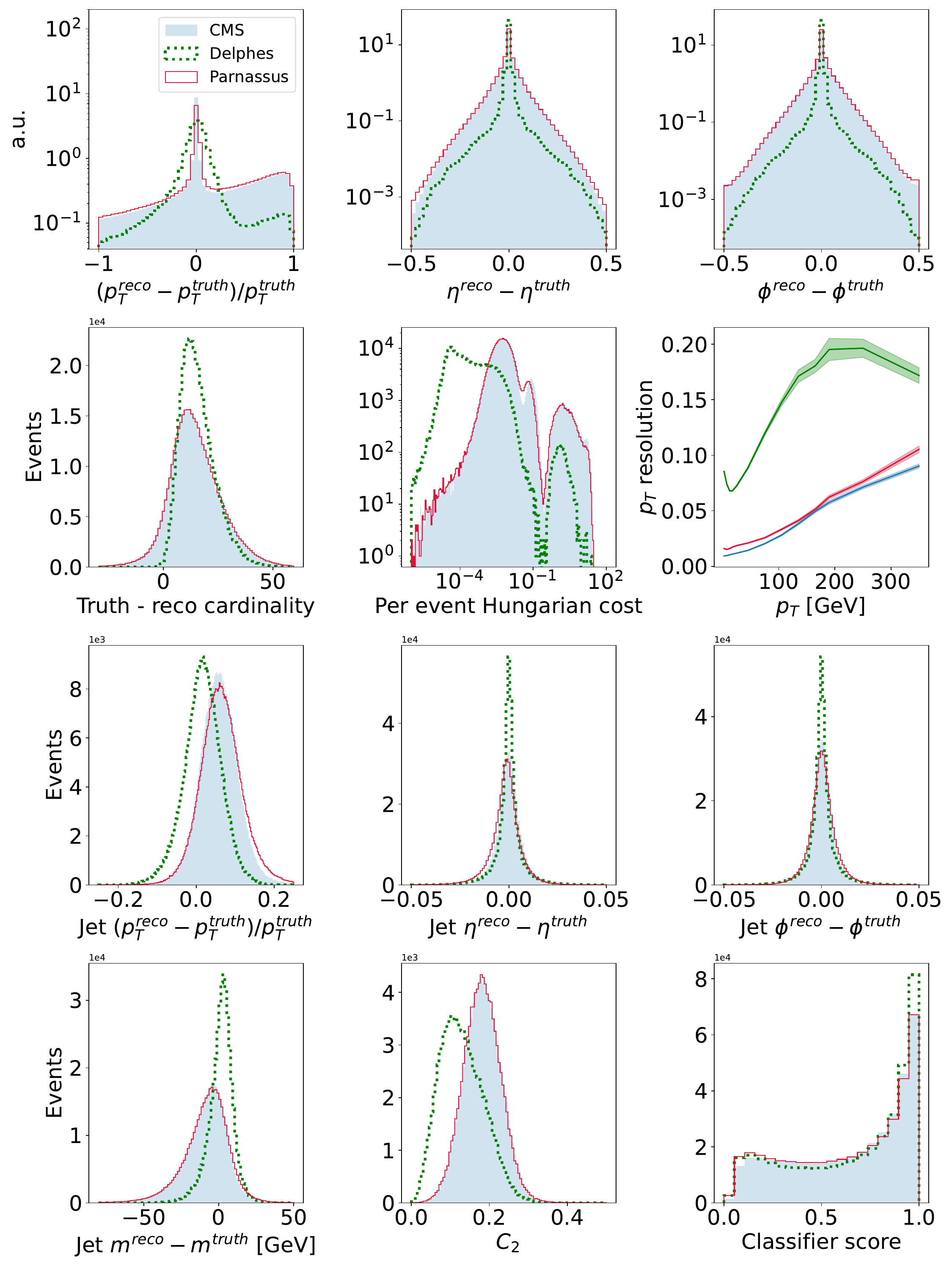}};
    \draw node [black] at (-6., 10.) {{\LARGE (a)}};
    \draw node [black] at (-0.8, 10.) {{\LARGE (b)}};
    \draw node [black] at (4.5, 10.) {{\LARGE (c)}};

    \draw node [black] at (-6., 10.-5.4) {{\LARGE (d)}};
    \draw node [black] at (-0.8, 10.-5.4) {{\LARGE (e)}};
    \draw node [black] at (4.5, 10.-5.4) {{\LARGE (f)}};

    \draw node [black] at (-6., 10.-5.4-5.4) {{\LARGE (g)}};
    \draw node [black] at (-0.8, 10.-5.4-5.4) {{\LARGE (h)}};
    \draw node [black] at (4.5, 10.-5.4-5.4) {{\LARGE (i)}};

    \draw node [black] at (-6., 10.-5.4-5.4-5.4) {{\LARGE (j)}};
    \draw node [black] at (-0.8, 10.-5.4-5.4-5.4) {{\LARGE (k)}};
    \draw node [black] at (4.5+2.6, 10.-5.4-5.4-5.4) {{\LARGE (l)}};
    
  \end{tikzpicture}
    \caption{In-distribution evaluation - distributions for jets that are statistically identical to the ones in the training dataset. (a) - (f) are constituent level and (g) - (l) are jet-level. See the text for variable definitions.} 
    \label{fig:indistribution}
\end{figure*}

We present two types of results. First, we examine how well the \ours model performs in the same regime as it was trained. For this purpose, we use a hold-out test set that is statistically identical to the training data. Figure~\ref{fig:indistribution} shows the performance on this test set for both local, particle-level properties as well as global, jet-level properties. In all cases, \ours agrees well with the CMS full simulation and reconstruction while there are clear challenges for \textsc{Delphes}. In particular, \textsc{Delphes} predicts a worse $p_T$ resolution and underestimates the angular resolution of individual constituents. This is true across particle $p_T$ as shown in Fig.~\ref{fig:indistribution}(f), which presents the standard deviation of a Gaussian fit around the peak of histograms like Fig.~\ref{fig:indistribution}(a) in bins of truth $p_T$.
In Fig.~\ref{fig:indistribution}(d), \textsc{Delphes} shows less variability in the number of PFCs relative to the number of true particles. Since it is not able to generate fake particles, \textsc{Delphes} almost always predicts fewer PFCs than truth particles. All of these features are captured by \ours.

At the jet level, we consider the results prior to applying a jet calibration. This means that the average $p_T$ response is not zero. In practice, one could apply the publicly available CMS jet calibrations~\cite{CMS:2016lmd} or a separate neural network could be derived to correct the average scale. Since \textsc{Delphes} essentially does not change the scale of jets (aside from losing some energy from certain particles), there is a clear shift in means for Fig.~\ref{fig:indistribution}(g). The jet angular resolution predicted by \textsc{Delphes} is too small while being well-approximated by \ours. The spectra for the jet substructure quantities are also modeled well by \ours while there are qualitative distortions in \textsc{Delphes}.

\begin{figure*}
    \centering
    \begin{tikzpicture}
    \path (0,0) node [red]                    {A};
     \node [inner sep=0] (image) {\includegraphics[width=0.9\textwidth]{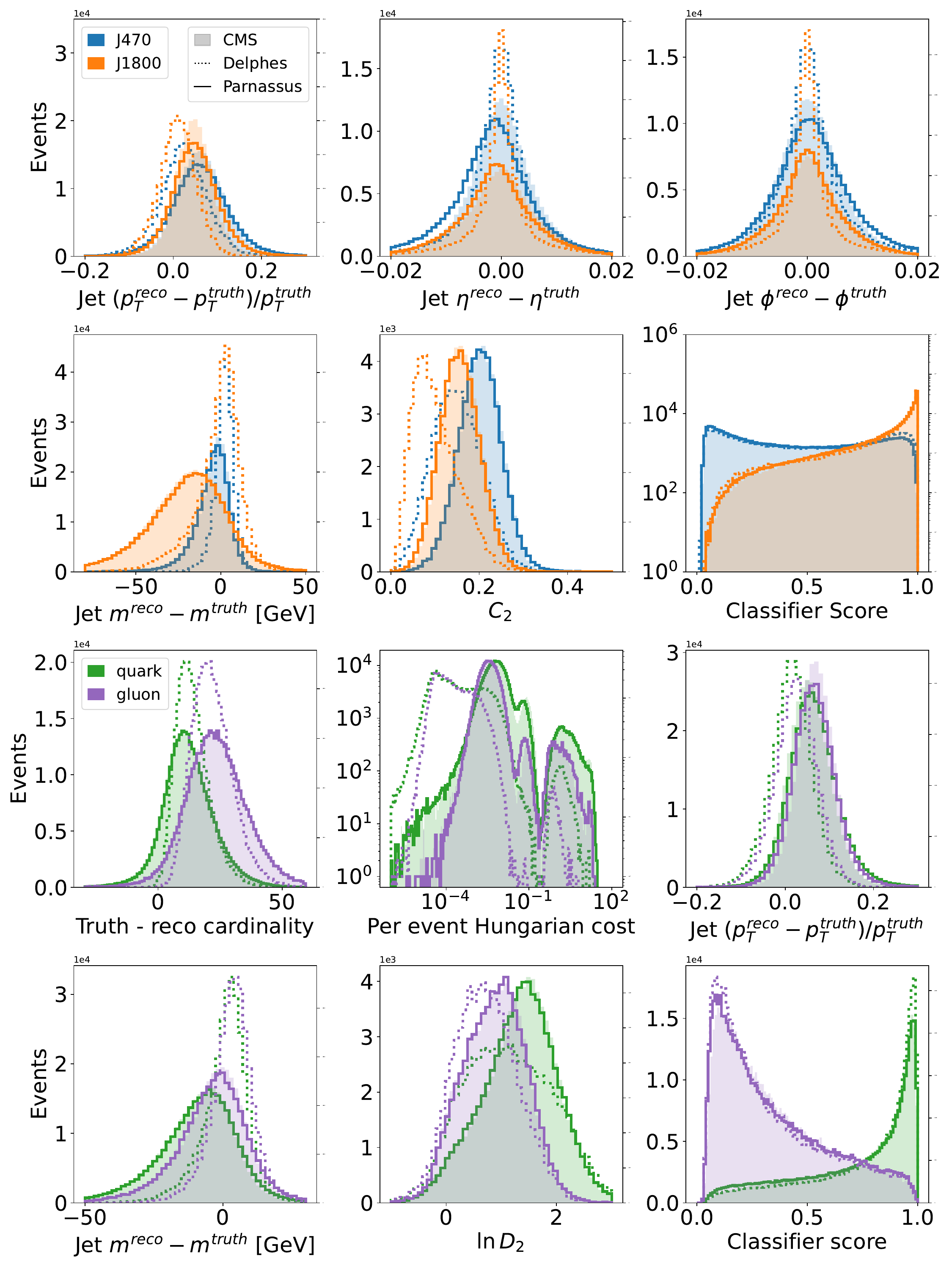}};
    \draw node [black] at (-6.2, 10.-1) {{\LARGE (a)}};
    \draw node [black] at (-0.9, 10.-0.2) {{\LARGE (b)}};
    \draw node [black] at (4.2, 10.-0.2) {{\LARGE (c)}};

    \draw node [black] at (-6.2, 10.-5.4-0.2) {{\LARGE (d)}};
    \draw node [black] at (-0.8+2.5, 10.-5.4-0.2) {{\LARGE (e)}};
    \draw node [black] at (4.5+2.5, 10.-5.4-0.2) {{\LARGE (f)}};

    \draw node [black] at (-6.2, 10.-5.4-5.4-1) {{\LARGE (g)}};
    \draw node [black] at (-0.8+2.5, 10.-5.4-5.4-0.1) {{\LARGE (h)}};
    \draw node [black] at (4.2, 10.-5.4-5.4-0.1) {{\LARGE (i)}};

    \draw node [black] at (-6.2, 10.-5.4-5.4-5.4) {{\LARGE (j)}};
    \draw node [black] at (-0.9, 10.-5.4-5.4-5.4) {{\LARGE (k)}};
    \draw node [black] at (4.5+1.2, 10.-5.4-5.4-5.4) {{\LARGE (l)}};
    
  \end{tikzpicture}
    \caption{Out-of-distribution evaluation. (a) - (f) are for lower and higher parton $p_T$ than used in training and (g) - (l) are for the training $p_T$ range, but split by quark and gluon jets. See the text for variable definitions.}
    \label{fig:outdistribution}
\end{figure*}

While it is encouraging that \ours performs so well in Fig.~\ref{fig:indistribution} relative to \textsc{Delphes}, it is also not surprising given that it was automatically tuned on the (statistically) same dataset while \textsc{Delphes} was tuned generically to CMS using per-object performance information. As a second set of results, we thus examine the performance of \ours on data distributions that it did not encounter during training. A set of plots for this out-of-distribution validation is shown in Fig.~\ref{fig:outdistribution}. 

The upper six plots of Fig.~\ref{fig:outdistribution} show the performance at lower and higher $p_T$ than used in the training. The jet datasets are split by parton $p_T$ and the range between 800 GeV and 1400 GeV was used for training. To show the extrapolation performance of the model, we examine the 470-600 GeV (J470) and 1800-$\infty$ GeV (J1800) datasets. There is some overlap in jet $p_T$ spectra between these datasets and the training one, but they also include much lower and higher momenta as well. The main reason to expect that \ours can extrapolate to these datasets is that while the ranges of jet $p_T$ are mostly orthogonal, large overlaps remain in the distributions of particle features; furthermore, we give the model the full truth spectrum. Figures~\ref{fig:outdistribution} (a) - (c) show that the response of the jet kinematic features are well reproduced by \ours. \textsc{Delphes} shows the same underestimation of the jet angular resolution present in Fig.~\ref{fig:indistribution}. The performance for jet substructure continues to be excellent for \ours despite large shifts in the spectra and resolutions with jet $p_T$. 

The lower six plots of Fig.~\ref{fig:outdistribution} demonstrate the performance for jets of different origins. The \ours model was not given parton labels during training, but we can examine post-hoc the performance on jets separated into quark- initiated and gluon-initiated jets. There is a fundamental ambiguity about the labeling of jets as originating from a quark or a gluon, but it is well-known (e.g. Ref.~\cite{ATLAS:2020cli,CMS:2016lmd}) that the response of jets depends on their origin, due to the differences in constituent multiplicity and (relative) energy spectra. Despite these differences, \ours is able to accurately match the CMS simulation and reconstruction. This is true for the slightly higher jet $p_T$ response for gluon jets relative to quark jets as well as for the more dramatic differences present in the other observables. The accurate description of a state-of-the-art quark/gluon jet tagger (Fig.~\ref{fig:outdistribution}(l)) shows that the accuracy of \ours is valid also for complex observables. The goal is that whatever observables are used for an analysis, \ours should be able to provide an accurate stand-in for the full simulation and reconstruction chain.

\section{Conclusion and Outlook}
\label{sec:conclusions}

We have presented \ours, a paradigm for automatically constructing a surrogate model for detector simulation and reconstruction. This framework is based on a deep generative model that takes as input a point cloud (particles impinging on a detector) and outputs another point cloud (reconstructed particle objects). While we have used the CMS detector as an example, this approach should be applicable to a variety of past, current, and future detectors in particle physics.

The long-term vision for \ours is that it will solve the critical computational challenges faced by large experiments and provide a common development framework for experimentalists and phenomenologists alike. The detector simulation and reconstruction of synthetic data within experiments is often an $O(1)$ fraction of all computing and thus if we can speed it up and make it compatible with hardware accelerators (e.g. GPUs), then this could no longer be a bottleneck for data analysis. Since \ours is portable, it would be simple for experiments to share their tuned model so that anyone recasting or projecting can use accurate simulations. As \ours is not based on any particle physics-specific software, it has the potential to lower the barrier of entry into particle physics.

While we have made a significant step towards this long-term vision, there are a number of milestones required to achieve the final product. Having focused here on jets, we need to consider entire events next and include additional output features such as particle type. Ensuring and quantifying universality will require demonstrating that the process is valid across a number of physics processes. To enable widespread use, \ours will require a convenient and configurable user interface like that of \textsc{Delphes}. Lastly, we will need to work with experiments to create models and facilitate the sharing of these models with the broader community. We believe that the progress presented in this paper serves as a proof-of-concept for \ours, establishing a new paradigm for fast, accurate, analysis-ready synthetic data in particle physics.

\section*{Data and Code Availability}

The code for this paper can be found at \url{https://github.com/parnassus-hep/cms-flow}. The postprocessed CMS dataset and the generated \textsc{Delphes} dataset can be found at Zenodo at \url{https://zenodo.org/records/11389651}.

\section*{Acknowledgments}
We thank Yaron Lipman and Nilotpal Kakati for fruitful discussions.  VM and BN are supported by the U.S. Department of Energy (DOE), Office of Science under contract DE-AC02-05CH11231.
EG, ED, DK, and NS are supported by the BSF-NSF grant 2028 and the ISF Research Center 494.

\appendix

\bibliography{HEPML,other,fastml}
\bibliographystyle{apsrev4-1}

\clearpage

\end{document}